\documentclass[sigconf,natbib=false]{acmart}
\setcopyright{none}
\renewcommand\footnotetextcopyrightpermission[1]{} 

\usepackage{graphicx}
\usepackage{subcaption}

\usepackage{tabularx}

\RequirePackage[
  datamodel=acmdatamodel,
  style=acmnumeric,
  ]{biblatex}

\begin{document}

\title{Graph Learning for Parameter Prediction of Quantum Approximate Optimization Algorithm}
\author{Zhiding Liang$^{*}$\ \   Gang Liu$^*$\ \  Zheyuan Liu$^*$\ \  Jinglei Cheng$^\P$\ \  Tianyi Hao$^\dag$\ \    Kecheng Liu$^*$\ \    Hang Ren$^\ddag$\ \   Zhixin Song$^\S$\ \   Ji Liu$^\|$\ \    Fanny Ye$^*$\ \    Yiyu Shi$^*$}
\affiliation{
  \institution{$*$ University of Notre Dame, $\P$ Purdue University, $\dag$ University of Wisconsin - Madison, $\ddag$ University of California, Berkeley, $\S$ Georgia Institute of Technology, $\|$ Argonne National Laboratory}
  \country{}
}

\begin{abstract}
In recent years, quantum computing has emerged as a transformative force in the field of combinatorial optimization, offering novel approaches to tackling complex problems that have long challenged classical computational methods. Among these, the Quantum Approximate Optimization Algorithm (QAOA) stands out for its potential to efficiently solve the Max-Cut problem, a quintessential example of combinatorial optimization. However, practical application faces challenges due to current limitations on quantum computational resource. Our work optimizes QAOA initialization, using Graph Neural Networks (GNN) as a warm-start technique. This sacrifice affordable computational resource on classical computer to reduce quantum computational resource overhead, enhancing QAOA's effectiveness. Experiments with various GNN architectures demonstrate the adaptability and stability of our framework, highlighting the synergy between quantum algorithms and machine learning. Our findings show GNN's potential in improving QAOA performance, opening new avenues for hybrid quantum-classical approaches in quantum computing and contributing to practical applications.
\end{abstract}





\maketitle

\section{Introduction}
In the burgeoning field of quantum computing, particularly within the realm of Noisy Intermediate-Scale Quantum (NISQ) devices~\cite{Preskill2018NISQ}, Variational Quantum Algorithms (VQAs) have emerged as a promising avenue for harnessing quantum advantages in the near term. NISQ devices, characterized by their limited number of qubits and inherent noise, present a unique computational landscape. Despite these constraints, they offer a practical platform for early quantum computations~\cite{wang2022torchquantum, wang2023robuststate, smith2022timestitch, tan2023hyqsat, zhang2024qplacer, zhang2023oneq}. The Quantum Approximate Optimization Algorithm (QAOA) is a prime example of algorithms tailored for NISQ machines~\cite{guerreschi2019qaoa, he2023alignment,lykov2023fast}, capitalizing on their capabilities to address complex combinatorial optimization problems like the Max-Cut problem. These NP-Hard problems are computationally challenging but hold immense practical significance in fields such as network design, data clustering, and circuit layout designs.

NISQ devices, though not yet capable of fully error-corrected quantum computations, still mark a significant step in the evolution of quantum technology. Their current limitations include shorter coherence times and higher error rates, which necessitate the development of specialized algorithms like VQAs that are resilient to these issues. The interplay between the hardware constraints of NISQ devices and the algorithmic ingenuity of VQAs represents a critical area of research in quantum computing. This synergy is at the heart of current efforts to unlock the potential of quantum computations in solving real-world problems~\cite{cheng2020accqoc, liang2023spacepulse, kashif2024hqnet, liang2024napa}, despite the nascent stage of quantum technology. The core of VQAs, and by extension, QAOA, revolves around the use of Parameterized Quantum Circuits (PQCs)~\cite{wang2022quantumnat, ravi2022cafqa, chu2023qdoor}, which function akin to quantum neural networks. The efficacy of these algorithms is deeply intertwined with their parameter initialization and optimization strategies, a critical aspect given the complex optimization landscapes characterized by issues like barren plateaus and local minima. Efficient initialization methods are particularly crucial in ensuring that the algorithm commences in close proximity to a potential solution within the parameter space, thereby facilitating more effective optimization~\cite{Google2021QAOA, liang2023hybrid}.

Recent trends in quantum computing have seen an intriguing amalgamation of classical and quantum learning architectures. Notably, the use of Graph Neural Networks (GNN) for solving combinatorial optimization problems~\cite{schuetz2022combinatorial} has demonstrated promising results. GNNs, leveraging their ability to directly process graph structures, have shown remarkable success in diverse applications ranging from social network analysis to biological network modeling~\cite{scarselli2008graph, zhou2020graph}. This adaptability makes them particularly suited for quantum computing tasks, where many problems can be naturally represented as graphs. In this paper, we delve into this hybridization by exploring the use of GNNs for the initialization of QAOA parameters. We posit that the integration of GNNs with quantum algorithms can significantly enhance the initialization process, particularly for complex problems like Max-Cut. This innovative approach opens up new avenues for leveraging the strengths of both machine learning and quantum computing to tackle some of the most challenging problems in computational science.

This paper explores the integration of Graph Neural Networks (GNN) with QAOA, aiming to enhance parameter initialization and extend the technique to weighted problems. Our focus is on harnessing the synergy between quantum computing and machine learning to tackle complex optimization tasks efficiently. Our main contributions are as follows:
\begin{itemize}
    \item We introduce a novel initialization method for QAOA parameters using Graph Neural Networks (GNN). This approach leverages the strengths of both quantum computing and machine learning. Our framework reduces the quantum resource overhead, making it more feasible for implementation on near-term quantum devices.
    \item We provide comprehensive benchmarking for different GNNs and analyze the suitable GNN for the QAOA case usage.
\end{itemize}

This paper explores the integration of GNN with QAOA for Max-Cut problems, starting with methodology, followed by experiments to validate our approach. We discuss data quality improvements, explore GNN architectures, and their impacts on QAOA. The paper concludes with discussions on outcomes and future work on AI meeting quantum computing.
\begin{figure}[h]
\centering
\includegraphics[width=\columnwidth]{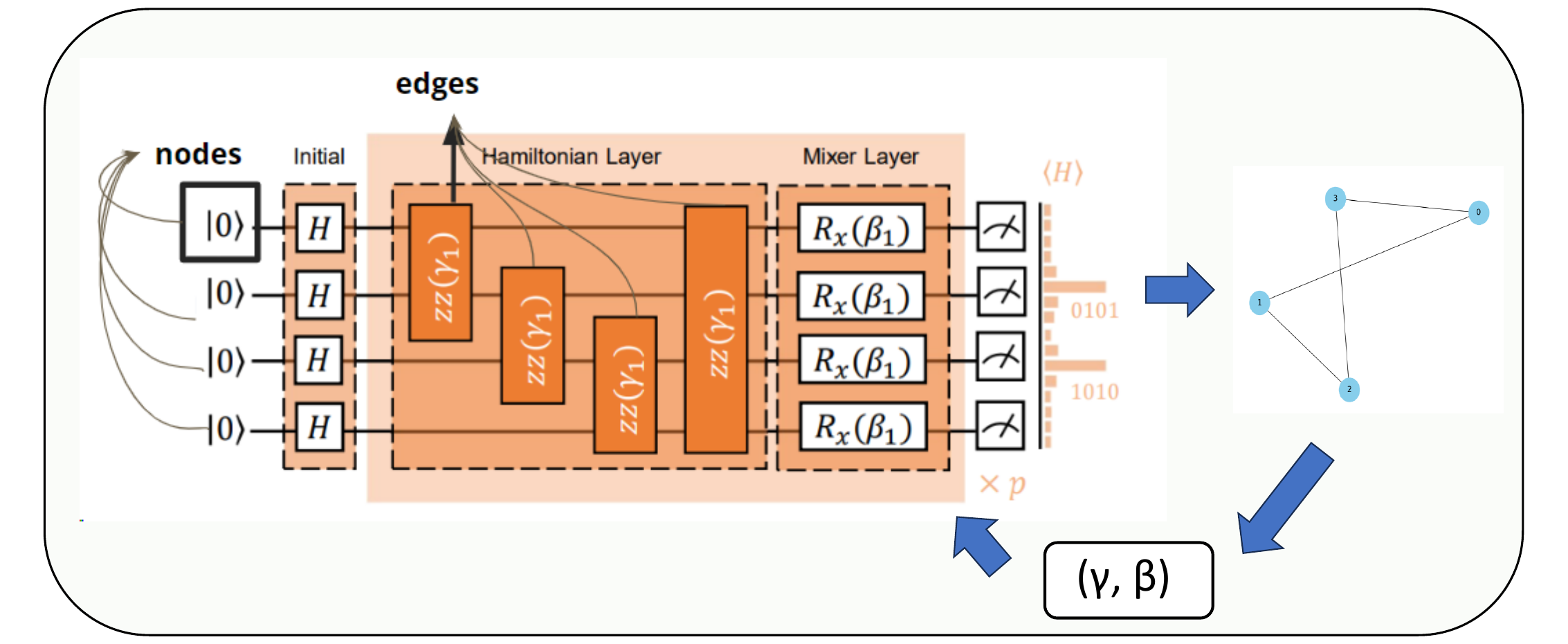}
\caption{The overview of the framework that uses the GNN to do the parameter ``warm-start''.}
\label{fig:overview}
\end{figure}

\section{Motivation}
In our quest to harness the full potential of quantum computing, particularly for the Quantum Approximate Optimization Algorithm (QAOA), we recognize the power of Graph Neural Networks (GNN) as a pivotal tool. The current era of Noisy Intermediate-Scale Quantum (NISQ) devices presents unique challenges. Not only are quantum computing resources inherently limited on each device, but access to real quantum computers is also not readily available for most researchers. This limitation is compounded by the fact that access typically requires reliance on platforms provided by large corporations, leading to long wait times and substantial costs.

Moreover, the aspiration for quantum computers is not merely to replace classical computers but to complement them, creating a synergy that tackles problems previously intractable by classical computation alone. This realization has spurred a growing interest in hybrid classical-quantum algorithms in recent years. Inspired by this trend, our work aims to leverage the synergy between GNN and QAOA. We propose using affordable classical computational resources to simulate and identify optimal initial parameters for QAOA, as illustrated in Figure \ref{fig:overview}. This strategy is not just about overcoming the hardware limitations of NISQ devices; it also represents a paradigm shift in how we approach quantum computing, making it more accessible and less financially burdensome for researchers. Our approach seeks to mitigate the constraints of NISQ devices, such as their limited qubit coherence and error rates, by optimizing the algorithmic efficiency on classical computers before execution on quantum hardware. This method promises to enable the QAOA to achieve convergence with fewer iterations on quantum computers, thereby enhancing its practicality and effectiveness. In doing so, we aim to unlock new potentials in NISQ devices, pushing the boundaries of what can be achieved in this nascent field.

Additionally, this approach has the potential to democratize access to quantum computing. By reducing the dependency on expensive quantum computing platforms and maximizing the preparatory work that can be done on classical systems, we open the door for a wider range of researchers to contribute to and benefit from advances in quantum computing. Our work not only addresses the technical challenges posed by the current state of quantum computing but also aligns with a broader vision of making quantum research more inclusive and sustainable. This endeavor could pave the way for groundbreaking advancements in the field, bringing the promise of quantum computing closer to realization in a variety of real-world applications.

\section{Methodologoy}
Since no open-source dataset is available for our experiments, the first step of our methodology is data preprocessing. Then, in the later part, we will elaborate on model preparation and the detailed structure of each benchmark. 
\subsection{Data Processing: } 
\begin{figure}[H]
    \centering
    \begin{subfigure}[b]{0.45\columnwidth}
        \includegraphics[width=\textwidth]{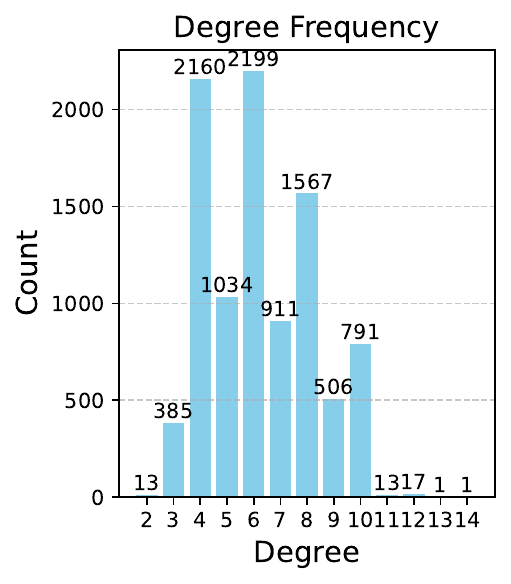}
        \caption{Degree frequency of target dataset.}
        \label{fig:degree_dis}
    \end{subfigure}
    \hfill
    \begin{subfigure}[b]{0.45\columnwidth}
        \includegraphics[width=\textwidth]{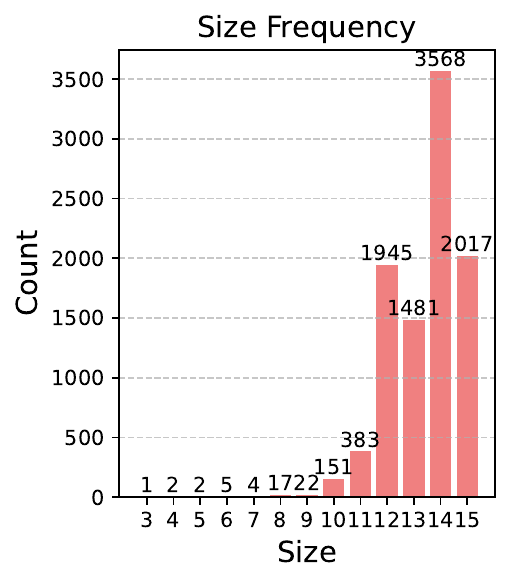}
        \caption{Graph size frequency of target dataset.}
        \label{fig:graph_dis}
    \end{subfigure}
    \caption{Degree and graph size distributions of the target dataset.}
    \label{fig:distributions}
\end{figure}

We generate synthetic regular graphs comprising 9598 instances and simulate the parameters $\gamma$ and $\beta$ for the QAOA algorithm.
These graphs vary in size, with nodes ranging from 2 to 15. The degree distribution of these graphs is captured in Figures \ref{fig:degree_dis} and \ref{fig:graph_dis}, illustrating that most graphs have a degree distribution between 2 to 14 and node numbers primarily distributed from 3 to 15. Each graph is stored in a text file, which is then inputted into the QAOA algorithm. The algorithm starts with randomly initialized values of $\gamma$ and $\beta$, and then undergoes a process of optimization over 500 iterations. This optimization seeks to refine the values of $\gamma$ and $\beta$, although the final values are optimized, they may not necessarily represent the absolute optimal parameters. 
In addition to determining the values of $\gamma$ and $\beta$, the QAOA algorithm outputs solutions for the Max-Cut problem. It also provides an approximation ratio (AR) for these solutions compared to the optimal solutions derived from a brute-force search approach.
We compute node degrees and one-hot encoding of node IDs as node features. The final output is an organized list comprising the graph structures along with important metadata like approximate ratio and values for the best cuts. This detailed dataset is then ready for further analysis and application in solving the Max-Cut problem using graph neural networks.

\subsection{Model Preparation: } 
In the pursuit of advancing the QAOA's application to the max-cut problem, our approach involves the integration of a multifaceted GNN-based prediction model. This model leverages a range of architectures—Graph Isomorphism Networks (GIN), Graph Attention Networks (GAT), Graph Convolutional Networks (GCN), and GraphSAGE—each selected for its unique capabilities in capturing the nuances of graph structure and node relationships. The GIN backbone anchors our model by encoding global graph properties, while GAT introduces an attention mechanism for nuanced node interaction, GCN simplifies neighborhood aggregation, and GraphSAGE optimizes for expansive neighborhood sampling.

Specifically, GNNs perform inference on data represented as graphs. These models leverage the intrinsic graph structure and node features to learn a vector representation for each node or for the entire graph. The core idea of GNNs is to update the representation of each node by aggregating feature information from its local neighborhood, which often includes the node's immediate neighbors. Through iterative aggregation and combination of these features, GNNs are able to capture the topological structure of the graph within a node's representation. Each iteration or layer in a GNN typically increases the receptive field of a node, allowing the incorporation of information from a wider neighborhood in the graph. 
Given the center node $v$ and the neighbor nodes $u \in \mathcal{N}(v)$ the representation of the node $h$ at the $k-1$ layer, the message passing or update function of GCN is shown as follows.
\begin{align}
a^{(k)}_v &= \text{AGGREGATE}^{(k)}\left( \left\{ h^{(k-1)}_u : u \in \mathcal{N}(v) \right\} \right), \\
h^{(k)}_v &= \text{COMBINE}^{(k)}\left( h^{(k-1)}_v, a^{(k)}_v \right),
\end{align}
where $\text{AGGREGATE}$ and $\text{COMBINE}$ are two functions crucial to the performance of GNN. The $\text{AGGREGATE}$ in GraphSAGE is:
\begin{equation}
    a^{(k)}_v = \text{MAX} \left( \left\{ \text{ReLU}(W \cdot h^{(k-1)}_u), \forall u \in \mathcal{N}(v) \right\} \right),
\end{equation}
where $W$ is the learnable model parameter. $\textbf{MAX}$ represents max-pooling in an element-wise. The Combination function in GraphSAGE could be:
\begin{equation}
    h^{(k)}_v = W [ h^{(k-1)}_v, a^{(k)}_v ].
\end{equation}

The original proposed GCN is analyzed with spectral graph convolutions and defined as follows.
\begin{equation}
    H^{(l+1)} = \text{ReLU} \left( \widetilde{D}^{-\frac{1}{2}} \widetilde{A} \widetilde{D}^{-\frac{1}{2}} H^{(l)} W^{(l)} \right). \tag{2}
\end{equation}
Here $H^{(l+1)}$ denotes node representation for all nodes for the next $l+1$ layer, $W$ $\widetilde{A} = A + I_N$, is the adjacency matrix of the undirected graph $G$ with added self-connections. $I_N$ is the identity matrix, $\widetilde{D}_{ii} = \sum_j \widetilde{A}_{ij}$ and $W^{(l)}$ is a layer-specific trainable weight matrix. According to the message passing definition, GCN could be rewritten as follows by integrating the aggregation and combination functions.
\begin{equation}
    h^{(k)}_v = \text{ReLU}\left( W \cdot \text{MEAN} \left\{ h^{(k-1)}_u, \forall u \in \mathcal{N}(v) \cup \{v\} \right\} \right).
\end{equation}
The aggregation function in GAT is:
\begin{equation}
    a^{(k)}_v = \text{MEAN} \left( \left\{ \alpha_{vu} \cdot h^{(k-1)}_u, \forall u \in \mathcal{N}(v) \right\} \right),
\end{equation}
where $\alpha_{uv}$ is the attention coefficient computed as:
\begin{equation}
    \alpha_{vu} = \frac{\exp\left( \text{LeakyReLU}\left( a^T [W h_v \| W h_u] \right) \right)}{\sum_{k \in \mathcal{N}(v)} \exp\left( \text{LeakyReLU}\left( a^T [W h_v \| W h_k] \right) \right)}.
\end{equation}
The aggregation function of GIN is summation, then the MLP is used for combination as follows:
\begin{equation}
    h^{(k)}_v = \text{MLP}^{(k)}\left( \left(1 + \epsilon^{(k)}\right) \cdot h^{(k-1)}_v + \sum_{u \in \mathcal{N}(v)} h^{(k-1)}_u \right),
\end{equation}
where $\epsilon$ may be a learnable parameter, the readout function for graph-level tasks is defined atop the node representations as follows.
\begin{equation}
    h_G = \text{READOUT}(\{h^{(K)}_v | v \in G\}).
\end{equation}
$h_G$ is the graph representation, which we could use with an MLP for prediction.

Our GNN predictor ingests graph data, including nodes and edges, augmented by additional features like node degrees and edge weights. The graph encoder plays a pivotal role, compressing node features into a lower-dimensional space rich with both attribute and structural fidelity. Following a mean-pooling layer that amalgamates node embeddings, the prediction layer comes into play, forecasting critical graph-level properties such as the QAOA parameters $\gamma$ and $\beta$. This comprehensive setup facilitates a meticulous comparison of each GNN architecture's performance against a baseline of random parameter initialization. By doing so, we aim to discern the impact of initial embeddings on the convergence efficiency and the overall solution quality, providing valuable insights into the optimization process within the QAOA landscape.

\subsection{Improvement of Data Quality: } 

\begin{figure}[h]
\centering
\includegraphics[width=1.0\columnwidth]{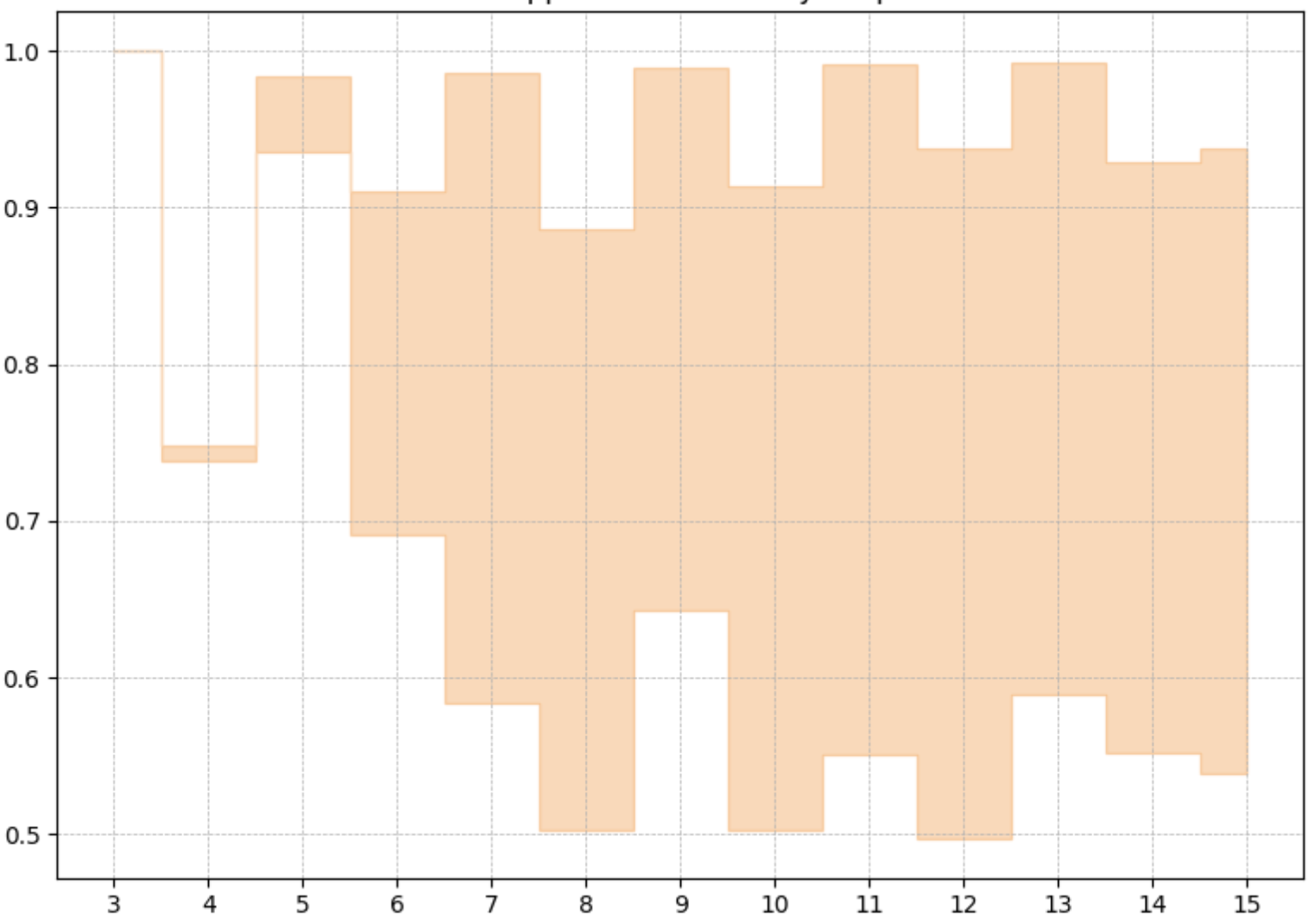}
\caption{Possible Approximation Ratio by Graph Size}
\label{fig:graph_size_quality}
\end{figure}

\begin{figure}[h!]
\centering
\includegraphics[width=1.0\columnwidth]{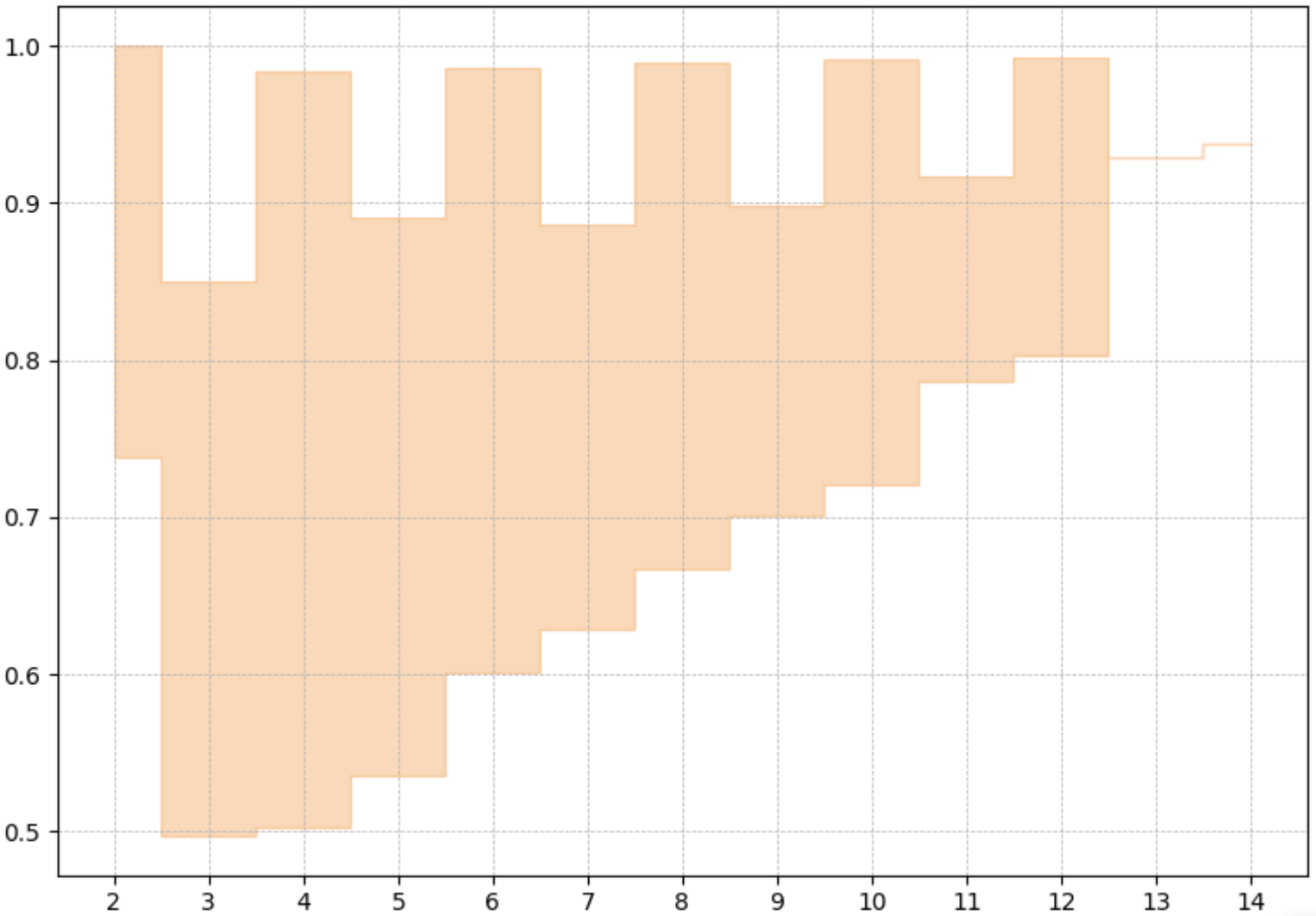}
\caption{Possible Approximation Ratio by Degree Number}
\label{fig:degree_num_quality}
\end{figure}

In our preliminary construction of a basic Graph Neural Network (GNN) for research purposes, we encountered a significant challenge regarding the quality of our dataset. This dataset comprises 9585 graphs, generated by applying random initialization as the parameter initialization method for the Quantum Approximate Optimization Algorithm (QAOA). Unfortunately, this approach has resulted in many instances of low-quality data within our dataset, which can be shown in Figure ~\ref{fig:graph_size_quality} and Figure ~\ref{fig:degree_num_quality}, where $x$ axis represents the graph size and degree number, y axis denotes the value interval, respectively. As it shown in the figure, we can observe that GNN training targets are not optimal, and have a relatively large gap compared to optimal solution in many cases, which may caused by noisy labels in the dataset. In particular, upon visualizing and analyzing this dataset, we observed that the approximation ratio for many groups of data was only around 50\%. The primary cause of this issue can be attributed to the inherently complex optimization landscape of the QAOA algorithm. Random initialization may lead the optimizer into regions where not even local optima exist. This scenario severely undermines the performance of the algorithm, preventing it from achieving the desired optimization results. Consequently, this section of our study is dedicated to exploring more effective ways to enhance data quality, aiming to improve the reliability of the GNN in handling these challenges.

\noindent\textbf{Fixed Parameter Conjecture: }In our research, we utilized a method from work~\cite{wurtz2021fixed} involving fixed parameter sets (angles) for QAOA, optimized based on tree subgraphs. These fixed parameters are suggested as a universal solution for regular graphs, potentially simplifying and accelerating QAOA by removing the need for individual graph parameter optimization. The findings show that these fixed angles yield performance close to optimal, providing strong heuristic evidence for their effectiveness in regular MaxCut graph QAOA. However, this method has a high computational cost. Despite searching in JPMorgan Chase's quantum team's open-source library~\cite{lykov2023fast}, a leader in the quantum algorithm industry, we only found results for regular graphs with degrees ranging from 3 to 11, which constitute about 6\% of our dataset, covering merely 587 graphs. This small subset's improvement, while notable, was too insignificant to substantially enhance the performance of our Graph Neural Network (GNN). The minimal impact of this method underscores the need for more comprehensive solutions to effectively improve GNN performance across the entire dataset.

\noindent\textbf{Selective Data Pruning: }In our advanced approach, the Selective Data Pruning (SDP) method was further refined to address the significant data quality issues in our quantum computing dataset. Recognizing that a substantial proportion of the dataset misdirected the GNN's learning, we initially set an approximation ratio threshold of 70\%, pruning data below this mark. However, this approach, while improving the overall data quality, led to a considerable loss of data. The reduced dataset size was insufficient for the GNN to adequately learn across the entire design space of the parameter space, hindering its ability to generalize and effectively model diverse quantum computing scenarios.

To address this issue, we introduced a selective rate, allowing us to fine-tune the balance between retaining and pruning poor-quality data. For instance, setting a selective rate of 70\% would mean preserving 70\% of the otherwise discarded data, while pruning the remaining 30\%. This nuanced approach provided a more balanced dataset, retaining enough data diversity for effective learning while still ensuring that the data quality was high enough to guide the GNN towards meaningful patterns and relationships. The introduction of the selective rate transformed the SDP method into a more dynamic and adaptable tool. It enabled us to iteratively refine our dataset, continuously assessing the impact of different selective rates on the GNN's performance. This iterative process was essential in identifying the optimal selective rate that strikes a balance between maintaining data quality and ensuring a robust dataset size for comprehensive learning and modeling.


\section{Experiments}

\begin{figure*}[!htbp]
\centering
\begin{subfigure}{0.245\textwidth}
            \includegraphics[width=\textwidth]
            {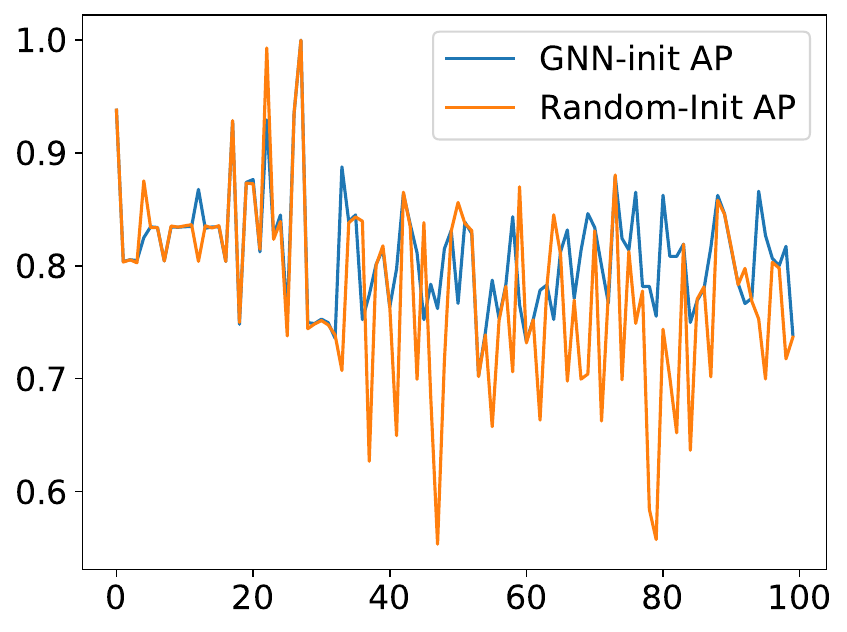}
            \subcaption{GAT}
        \end{subfigure}
        \begin{subfigure}{0.245\textwidth}
            \includegraphics[width=\textwidth]
            {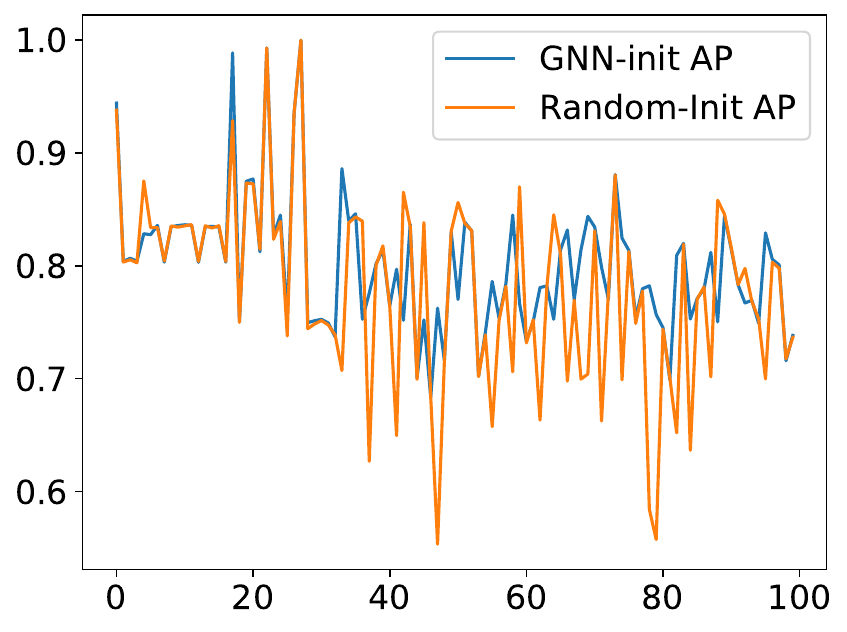}
            \subcaption{GCN}
        \end{subfigure}
        \begin{subfigure}{0.245\textwidth}
            \includegraphics[width=\textwidth]{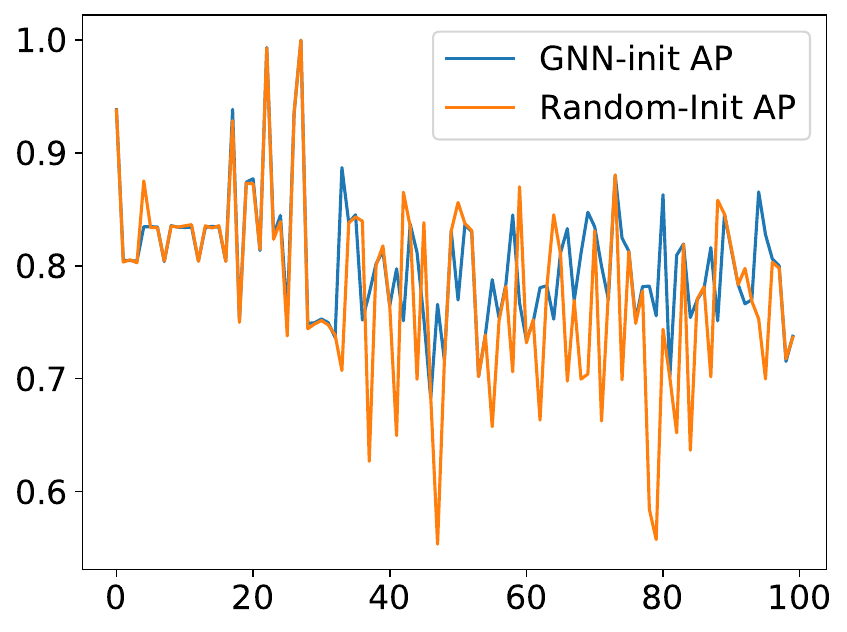}
            \subcaption{GIN}
        \end{subfigure}
        \begin{subfigure}{0.245\textwidth}
            \includegraphics[width=\textwidth]{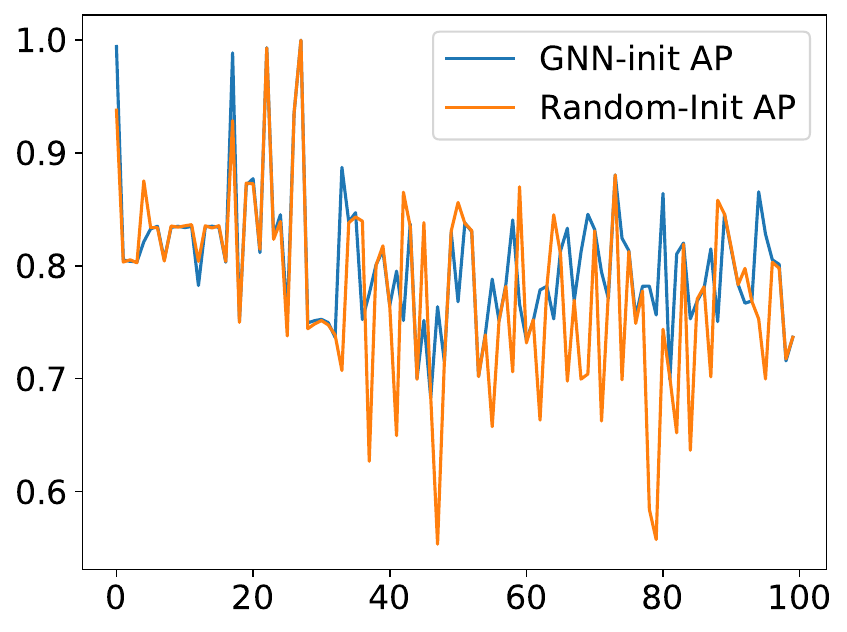}
            \subcaption{GraphSAGE}
        \end{subfigure}
\caption{
Comparison of Approximation Ratio between random initialisation and various GNN benchmarks. From left to right, the order is GAT, GCN, GIN and GraphSAGE.
}
\label{fig:approxi_ratio}
\end{figure*}

In this section, we conduct experiments to validate the effectiveness of graph benchmarks in solving QAOA problem under fixed parameters setting. 

\subsection{Experiment Setup}
\textbf{Dataset and Baseline Models.} We set aside 100 test graphs with different degrees and graph sizes to calculate the improvement in the approximation ratio achieved by different GNN-based QAOA initialisation. In particular, we focus on four GNN benchmarks: GCN, GAT, GIN and GraphSAGE and one baseline - random intialisation. The results for these 100 instances are presented in Figure \ref{fig:approxi_ratio}, where the orange line is the approximation ratio for random initialisation and the blue line the approximation ratio for various GNNs. The break down results are presented in Table ~\ref{tab:AP_result}.

\noindent\textbf{Implementation Details.}
We report the mean and standard deviation of the result across 100 test graphs. For each GNN model, we set the input dimension to 15, number of GNN layer to 2. Furthermore, we set embedding dimension to 32 and dropout ratio to 0.5 during training, which ensures that no single neuron becomes overly specialized to rely on specific features of the input data. This encourages the network to learn more robust features that generalize better to unseen data. We use Adam optimizzer to optimize the model and ReduceLROnPlateau as scheduler to monitor the training loss and reduces the learning rate when there is no improvements for a defined number of epochs. In particular, we set scheduler mode to min, factor to 5, patience to 5 and minimum learning rate to 1e-5. Lastly, we train each model for 100 epoches before examine it on test set.

\begin{table}[t]
    \centering
    \renewcommand*{\arraystretch}{2} 
    \setlength{\tabcolsep}{2.8pt} 
    \footnotesize
    \begin{tabular}{|c|c|c|c|c|} 
    \hline
Methods & GAT & GCN & GIN & GraphSAGE \\ \hline
 Average Improvement  & $3.28_{\pm 9.99}$ & $3.65_{\pm 10.17}$ & $3.66_{\pm 9.97}$ & $2.86_{\pm 10.01}$ \\ 
        \hline 
    \end{tabular}
    \caption{Average improvements of GNN benchmarks compared to random initialization.}
    \label{tab:AP_result}
\end{table}



\subsection{Result Analysis}
In our experiment comparing various Graph Neural Network (GNN) benchmarks for initializing the Quantum Approximate Optimization Algorithm (QAOA) against random initialization, the results showed varied performance across different GNN architectures. The baseline, random initialization, serves as a comparison point. Graph Attention Networks (GAT)~\cite{velivckovic2017graph} slightly outperformed compared to random initialization, with a score of $3.28_{\pm 9.99}$. Graph Convolutional Networks (GCN)~\cite{kipf2016semi} improved over the baseline, scoring $3.65_{\pm 10.17}$, indicating that convolutional approaches in GCN might be capturing useful features for QAOA, but not to a large extent. Graph Isomorphism Networks (GIN)~\cite{xu2018how} showed slightly better performance with $3.66_{\pm 9.97}$, potentially due to their ability to capture structural information in graphs. In contrast, GraphSAGE~\cite{hamilton2017inductive}, with a score of $2.86_{\pm 10.01}$, improved only a bit in terms of performance compared to random initialization, denoting that its inductive learning approach might not align well with QAOA initialization requirements. And from Figure \ref{fig:approxi_ratio}, we can see that across test graphs, the performance of GNN benchmarks is more stable than the random initialization approach. For example, the Graph Isomorphism Networks (GIN) show a relatively more stable performance with fewer instances where random initialization surpasses GNN initialization, suggesting a potentially more reliable performance. 
\section{Related Work}
\cite{egger2021warm} using Goemans-Williamson random rounding to warm-start recursive QAOA and shown a consistent improvement in the cut size for fully connected graphs with random weights for the MaxCut problem. This approach can also be applied to other random rounding schemes and optimization problems. And recent work~\cite{tate2023bridging} introduce a classical pre-processing step that initializes QAOA with a biased superposition of possible cuts in the graph, referred to as a warm-start. A close work~\cite{jain2022graph} use GCN and line graph neural network (LGNN) for warm start the QAOA. In their framework, GCN and LGNN update graph embeddings based on neighbors using a function $f\theta$, outputting probabilities for each node's cut side. 

\section{Conclusion}

In conclusion, this paper presents a significant step forward in the integration of Graph Neural Networks (GNN) with the Quantum Approximate Optimization Algorithm (QAOA), specifically targeting the Max-Cut problem. To show the potential of GNNs in enhancing the initialization process of QAOA parameters, we established a series of GNN benchmarks and test their performances in fixed parameter settings. Through extensive experiments and analyses, we have identified key areas for improvement and future research. Through extensive experiments and detailed analyses, we have uncovered several key areas that hold promise for further research and improvement. Our findings indicate that the integration of GNNs with QAOA can be significantly enhanced through refined data pruning strategies, adaptive learning rate mechanisms, and advanced network architectures. These improvements are crucial for handling the intricacies of quantum data and for achieving optimal algorithmic performance. Additionally, our work opens up new avenues for exploring the synergy between classical machine learning techniques and quantum algorithms. It suggests that similar approaches could be effectively applied to other quantum algorithms and problems, potentially leading to breakthroughs in quantum computation efficiency and effectiveness. Looking ahead, we aim to continue this line of research by exploring more sophisticated GNN models and delving deeper into the quantum-classical interface. The goal is to further improve the initialization process for QAOA and other quantum algorithms, thereby enhancing their applicability and performance in solving real-world problems. This work lays the foundation for future explorations in quantum computing and positions GNNs as a pivotal tool in this rapidly evolving field.

\section{Future Works and Limitations}
Here, we recognize some limitations of our project and raise some possible future improvements. Specifically, we categorize challenges into three main areas: problem definition, data quality, and the suitability of GNN structures for QAOA. Firstly, the existing models are primarily designed for unweighted graphs, leading to inconsistent performance on weighted graphs, which are more common in real-world scenarios. This limitation prevents the model's broader applicability. Secondly, data quality poses significant challenges as we mentioned in methodology. The models need datasets with higher noise levels to test robustness, but this can complicate training and lead to unreliable predictions. Also, there's a gap between the training targets of the GNN and the optimal solutions, possibly due to noisy data affecting the training accuracy. Lastly, the current GNN structures might not be fully optimized for QAOA requirements, potentially leading to inefficient initializations and suboptimal optimization results, as it shown in the result section where the performance of GNN benchmark is not significantly exceed the random initialisation baseline. To address these challenges, we need to develop more advanced GNN architectures that can effectively process weighted graphs and withstand noisy data. Secondly, it is imperative to further improve data preprocessing method to minimize the impact of noise and enhance the quality of training labels. Thirdly, it is also critical to redefine the problem to cover both weighted and unweighted graphs, which will be helpful in creating a more versatile and applicable model for quantum optimization tasks.

\section{Acknowledgment}
This material is supported by the DOE-SC Office of Advanced Scientific Computing Research AIDE-QC project under contract number DE-AC02-06CH11357.
\printbibliography
\end{document}